# Multimodal risk trajectories reveal heterogeneous paths to dementia

Zhiqi Lee[1], Haowen Li[1], Tao Liu[1, 2], Shiyuan Zhang[1], Bingjie Wang[1], Ruhan Bi[4], Jinzhao Fan[1], Yunkai Zhang[1], Bei Zhang[4], Zhuonan Wang[3, *], Lijun Bai[1, *]

1. Department of Biomedical Engineering, School of Life Science and Technology, The Key Laboratory of Biomedical Information Engineering Ministry of Education, Xi'an Jiaotong University
2. Medical Imaging Center of Xi'an QIN HUANG Hospital
3. PET-CT Center, the First Affiliated Hospital, Xi 'an Jiaotong University
4. Brain Health Institute, National Center for Mental Disorders, Shanghai Mental Health Center, Shanghai Jiao Tong University School of Medicine and School of Psychology

* Correspondence and requests for materials should be addressed to Lijun Bai (email: bailijun@xjtu.edu.cn) or Zhuonan Wang (email: wangzhuonan@xjtu.edu.cn).

**Abstract**

Dementia comprises biologically heterogeneous disorders, yet current risk assessment provides limited insight into how subtype-specific risk emerges and diverges before clinical diagnosis. We developed **NetMoint**, a multimodal framework integrating partially observed plasma proteomic, structural magnetic resonance imaging and cerebral haemodynamic phenotypes to predict individualized risks of Alzheimer's disease (AD), vascular dementia (VD) and frontotemporal dementia (FTD) across 1-, 5-, 10- and 20-year horizons. Among 104,120 UK Biobank participants free of dementia at baseline, NetMoint achieved mean area under the receiver operating characteristic curve (AUC) values of 0.937, 0.930 and 0.932 for AD, VD and FTD, respectively. The biological determinants of prediction shifted with time, from structural brain vulnerability at shorter horizons towards circulating molecular signatures at longer horizons, with distinct subtype-specific biological profiles. Multi-horizon risk profiling identified distinct temporal trajectories of dementia susceptibility. Among participants who subsequently developed AD, 0.7% followed a persistently very-high-risk trajectory, with predicted risk reaching 53.50% at 20 years, whereas 8.3% of those who developed FTD followed an increasing very-high-risk trajectory, reaching 67.17%. These high-risk trajectories were marked by distinct molecular signatures, with lower TGFB1 characterizing the AD group and higher NDRG1 the FTD group. In an independent ADNI-to-UK Biobank analysis, AD risk prediction remained informative after harmonization to 138 shared features, with an AUC of 0.741 at 20 years. Together, these findings establish a multimodal framework for trajectory-resolved dementia risk stratification, identifying small but high-risk populations within dementia subtypes and linking their divergent risk trajectories to distinct molecular signatures.

# Introduction

Accurate identification of individuals at high risk of dementia before substantial and potentially irreversible neurodegeneration has become an increasingly important clinical priority as disease-modifying therapies enter clinical practice [1, 2]. Dementia comprises biologically distinct disorders, including Alzheimer's disease (AD), vascular dementia (VD) and frontotemporal dementia (FTD), that often present with overlapping clinical manifestations [3-5]. Because their pathological processes begin years before clinical onset [1, 2], a key unmet need is to determine not only who is at risk and which subtype may develop, but also when that risk becomes clinically meaningful, to enable precision prevention, clinical-trial enrichment and early therapeutic intervention [6, 7].

Current approaches to dementia risk assessment draw primarily on clinical characteristics, structural neuroimaging and blood-based biomarkers [6, 8-10]. These measurements capture complementary aspects of the preclinical disease process but provide different biological windows. Structural MRI reflects neuroanatomical changes associated with neurodegeneration, which may become more evident as disease advances [6, 10]. Blood proteomics can detect molecular alterations years before clinical diagnosis and has shown promise for dementia risk prediction and subtype discrimination [11-14], although circulating proteins are also influenced by systemic processes and may not fully reflect pathology within the central nervous system [15-17]. Cerebral haemodynamic measures provide a further dimension, linking vascular dysfunction with impaired perfusion, metabolic homeostasis and clearance mechanisms implicated in neurodegeneration [18, 19]. These complementary signals raise the possibility that different biological systems may become informative at different stages of the preclinical course, making their integration potentially more informative than any individual modality [9, 17, 20].

Yet dementia risk is still largely framed as a static prediction problem. Most existing models estimate all-cause dementia or a single subtype, often optimized for a predefined prediction horizon [8, 11-13]. Such approaches can establish who is more likely to develop dementia but provide limited information about the trajectory through which risk accumulates, whether that trajectory differs between dementia subtypes, or whether individuals who ultimately receive the same diagnosis follow the same path before onset. This distinction may be clinically important. A small proportion of individuals may carry a disproportionately high future risk, whereas others with the same eventual diagnosis may remain at relatively low predicted risk until closer to clinical onset. Moreover, the biological signals that identify near-term susceptibility may not be the same as those that characterize risk decades before diagnosis. Whether dementia susceptibility therefore comprises distinct temporal trajectories, and whether the information supporting these trajectories changes with the prediction horizon, remains poorly understood.

A second challenge is practical. Participants in population studies rarely undergo identical combinations of proteomic and imaging assessments, and requiring complete measurements across modalities can substantially reduce usable sample size and limit clinical applicability [9, 10]. Artificial intelligence offers a means of integrating high-dimensional, nonlinear and complementary biological information [21-23], but existing multimodal approaches have largely focused on current disease classification or prediction at a fixed time point. Recent studies have demonstrated the value of multimodal AI for differential dementia diagnosis and of plasma proteomics for neurodegenerative disease classification and biomarker discovery [5, 11-14]. What remains less clear is whether multimodal integration can move beyond a single risk estimate to resolve the temporal structure of future dementia susceptibility and identify biologically distinct high-risk groups within the same clinical subtype.

Here, we developed NetMoint (Network-based Multi-Omics Integration), a multimodal artificial intelligence framework that integrates partially observed plasma proteomic, cerebral haemodynamic and structural MRI data to estimate individualized risks of AD, VD and FTD across 1-, 5-, 10- and 20-year horizons. Using 104,120 dementia-free participants from the UK Biobank followed for up to 20 years, we tested whether dementia susceptibility is distributed across distinct temporal trajectories and whether the biological information underlying prediction changes with the prediction horizon. We then examined whether these trajectories remain heterogeneous among individuals who ultimately develop the same dementia subtype, and whether molecular features can distinguish trajectory-defined high-risk groups. Finally, we evaluated the transportability of AD prediction in an independent ADNI cohort. By linking multimodal risk trajectories to their temporal and molecular determinants, this framework seeks to move dementia risk assessment beyond a single probability of future disease towards a more resolved characterization of who is at risk, how that risk evolves and which individuals may warrant closer biological surveillance.

## Results

### 1. Cohort characteristics and multimodal profiling of dementia vulnerability

We assembled a longitudinal cohort of 104,120 UK Biobank participants with baseline phenotypic, imaging, proteomic and clinical data (Fig. 1a). Participants had a mean age of 60.4 ± 8.8 years (range, 39.0–70.0 years), and 53.5% were women. During up to 20 years of follow-up, 1,124 participants developed incident all-cause dementia, including 627 with Alzheimer's disease (AD), 273 with vascular dementia (VD) and 109 with frontotemporal dementia (FTD); participants could receive more than one dementia subtype diagnosis. The all-cause dementia group also included individuals with other or unspecified dementia diagnoses, whereas the remaining 102,996 participants had no recorded dementia diagnosis and served as controls.

Participants who subsequently developed dementia already differed from controls across multiple clinical domains at baseline (Table 1). Across dementia subtypes, affected individuals were older and had greater burdens of cardiometabolic, vascular and psychiatric risk factors. Participants who later developed AD had higher prevalence of hypertension, hyperlipidaemia, diabetes, atrial fibrillation and depression, whereas those who developed VD showed the strongest enrichment of vascular risk factors, including hypertension, diabetes, atrial fibrillation, previous stroke and smoking. Individuals who developed FTD also showed higher metabolic and psychiatric comorbidity, but a less pronounced vascular burden.

Beyond conventional risk factors, participants who subsequently developed dementia already showed subtle cognitive differences years before diagnosis. Baseline fluid intelligence scores were lower across AD, VD and FTD groups compared with controls, accompanied by prolonged reaction times, indicating early alterations in cognitive function preceding clinically diagnosed dementia. Together, these findings demonstrate that incident dementia subtypes are preceded by distinct multimodal biological and cognitive signatures, providing a foundation for modelling individual ageing trajectories and disease vulnerability.

**Tabel 1 Baseline characteristics of UK Biobank participants included in the study**

| characteristics | Overall | Control | Incident AD | P value | Incident VD | P value | Incident FTD | P value |
|---|---|---|---|---|---|---|---|---|

| [subjects] | (N=104120) | (N=102996) | (N=627) | | (N=273) | | (N=109) | |
|---|---|---|---|---|---|---|---|---|
| Age, years | 55.7 [39.0-70.0] | 55.7 [39.0-70.0] | 64.9 [42.0-70.0] | <0.001 | 65.0 [42.0-70.0] | <0.001 | 62.0 [44.0-70.0] | <0.001 |
| Sex (female) | 55667 (53.5%) | 55229 (53.5%) | 338 (53.9%) | 0.87 | 99 (36.3%) | <0.001 | 47 (43.1%) | 0.038 |
| Enhanced PRS for AD | 0.0 [-3.4-4.9] | 0.0 [-3.4-4.6] | 1.2 [-1.7-4.9] | <0.001 | 0.6 [-1.2-3.6] | <0.001 | 0.2 [-1.7-1.8] | 0.376 |
| Stroke | 2482 (2.4%) | 2373 (2.3%) | 53 (8.5%) | <0.001 | 67 (24.5%) | <0.001 | 5 (4.6%) | 0.203 |
| Diabetes | 7577 (7.3%) | 7359 (7.1%) | 135 (21.5%) | <0.001 | 93 (34.1%) | <0.001 | 22 (20.2%) | <0.001 |
| Heart failure | 3384 (3.3%) | 3270 (3.2%) | 68 (10.8%) | <0.001 | 55 (20.1%) | <0.001 | 9 (8.3%) | 0.006 |
| Atrial fibrillation | 7132 (6.8%) | 6940 (6.7%) | 133 (21.2%) | <0.001 | 73 (26.7%) | <0.001 | 13 (11.9%) | 0.048 |
| Peripheral artery disease | 1481 (1.4%) | 1432 (1.4%) | 29 (4.6%) | <0.001 | 26 (9.5%) | <0.001 | 1 (0.9%) | 0.992 |
| Hypertension | 28981 (27.8%) | 28439 (27.5%) | 372 (59.3%) | <0.001 | 201 (73.6%) | <0.001 | 46 (42.2%) | <0.001 |
| Depression | 5778 (5.5%) | 5587 (5.4%) | 122 (19.5%) | <0.001 | 72 (26.4%) | <0.001 | 29 (26.6%) | <0.001 |
| Hyperlipidemia | 15035 (14.4%) | 14713 (14.3%) | 210 (33.5%) | <0.001 | 126 (46.2%) | <0.001 | 35 (32.1%) | <0.001 |
| Systolic blood pressure, mmHg | 138.1 [68.0-252.0] | 138.1 [68.0-252.0] | 146.9 [78.0-223.0] | <0.001 | 145.4 [78.0-220.0] | <0.001 | 141.6 [106.0-195.0] | 0.061 |
| Past tobacco smoking | 54737 (55.9%) | 54220 (55.8%) | 355 (59.8%) | 0.06 | 169 (66.8%) | <0.001 | 61 (59.2%) | 0.553 |
| Pairs matching time(s) | 112.3 [0.0-6510.0] | 111.6 [0.0-6510.0] | 189.3 [0.0-2764.0] | <0.001 | 185.1 [0.0-1432.0] | <0.001 | 170.3 [0.0-1811.0] | 0.017 |
| Fluid intelligence score | 6.3 [0.0-13.0] | 6.3 [0.0-13.0] | 5.0 [1.0-12.0] | <0.001 | 4.7 [1.0-9.0] | <0.001 | 4.6 [0.0-10.0] | <0.001 |

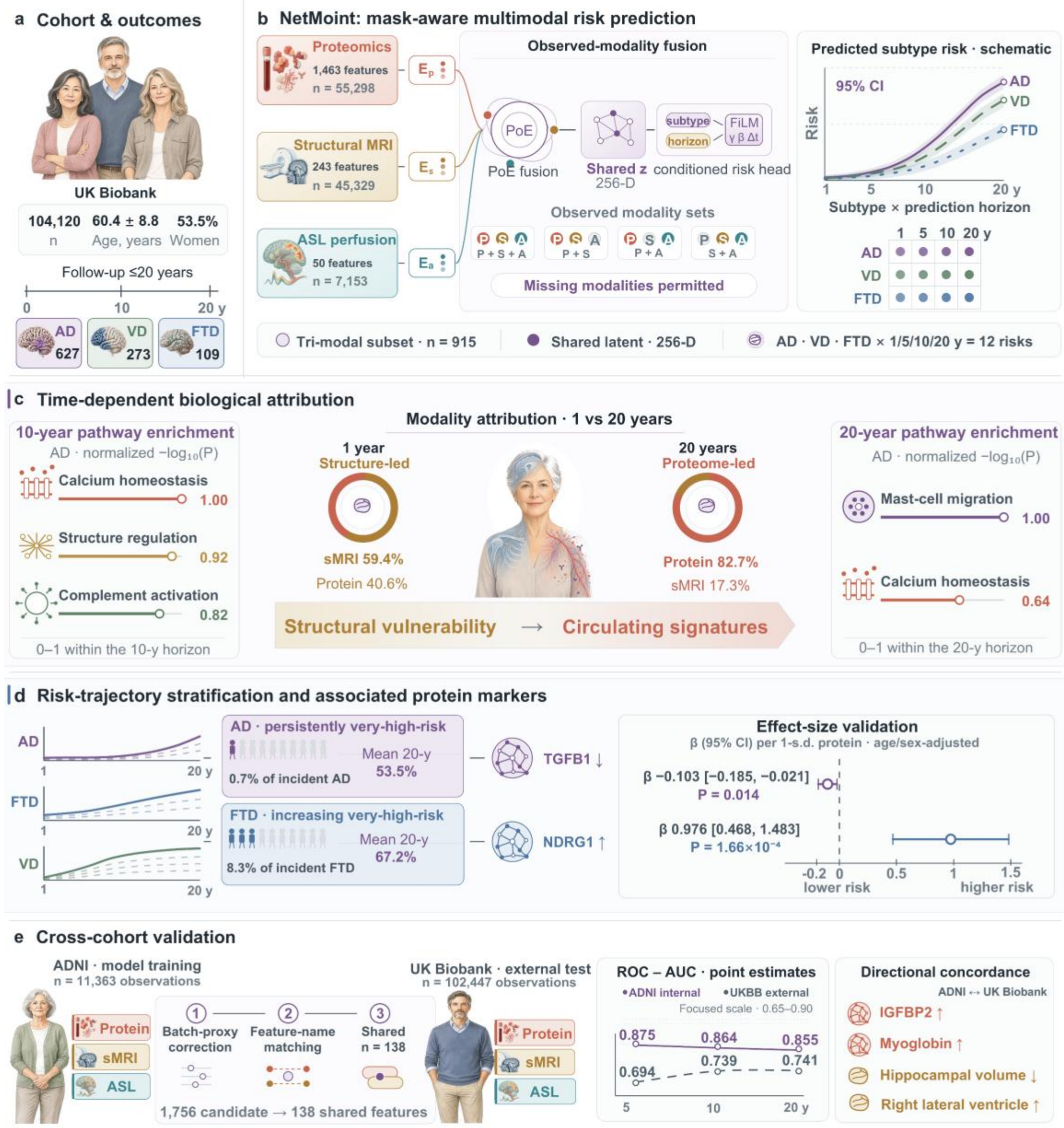

**Fig. 1 | Study design, NetMoint architecture and principal findings. a**, The study included 104,120 UK Biobank participants, among whom 627 developed Alzheimer's disease (AD), 273 vascular dementia (VD) and 109 frontotemporal dementia (FTD) during follow-up; subtype labels were not mutually exclusive. **b**, NetMoint integrated plasma proteomics, structural MRI and arterial spin labelling (ASL) MRI using modality-specific encoders and precision-weighted product-of-experts fusion to generate shared representations and 12 subtype-specific risk estimates across 1-, 5-, 10- and 20-year horizons. **c**, Feature attribution varied by prediction horizon, with structural MRI contributing most at 1 year and plasma proteomics at 20 years; pathway analyses identified calcium-homeostasis, structural-regulation, complement and mast-cell migration processes. **d**, Subtype-specific trajectories revealed small but highly enriched high-risk groups; the very-high-risk AD trajectory comprised only 0.7% of the cohort but had a mean predicted 20-year risk of 53.5%. TGFB1 was inversely associated with predicted AD risk, whereas NDRG1 showed a stronger association with predicted FTD risk at longer horizons. **e**, External validation in UK Biobank after training in the Alzheimer's Disease Neuroimaging Initiative yielded ROC–AUC values of 0.694, 0.739 and 0.741 at 5, 10 and 20 years, respectively, with concordant plasma and structural MRI signals; ASL findings remained exploratory because of sparse

events. Risk estimates represent model-derived probabilities and not observed cumulative incidence or biological disease progression. ADNI, Alzheimer's Disease Neuroimaging Initiative; ASL, arterial spin labelling; PoE, product of experts; sMRI, structural magnetic resonance imaging.

## 2. NetMoint enables subtype-specific dementia risk prediction across multiple timescales

We next developed NetMoint (Fig. 1b) to integrate heterogeneous multimodal phenotypes and estimate individualized risks of Alzheimer's disease (AD), vascular dementia (VD) and frontotemporal dementia (FTD) across 1-, 5-, 10- and 20-year horizons (Fig. 1). The model combined modality-specific representations with subtype-specific prediction heads and was applied to the 104,120 participants who were free of AD, VD and FTD at baseline. Multimodal inputs comprised 1,463 circulating proteomic features from 55,298 participants, 243 structural MRI features from 45,329 participants and 50 cerebral haemodynamic MRI features from 7,153 participants, with 915 participants having measurements across all three modalities.

NetMoint showed consistent discrimination across prediction horizons, with mean area under the receiver operating characteristic curve (AUC) values of 0.937 for AD, 0.930 for VD and 0.932 for FTD across the four horizons (Fig. 2a). Temperature scaling yielded well-calibrated predicted probabilities, which enabled risk stratification across the cohort. The separation between participants who subsequently developed dementia and controls became particularly pronounced at longer horizons: at 20 years, mean predicted risks were 25.7-fold higher for AD, 64.4-fold higher for VD and 135.2-fold higher for FTD among participants who subsequently developed the respective dementia subtype than among controls (Fig. 2b). These results established that multimodal baseline phenotypes could support accurate subtype-specific risk estimation over substantially different timescales.

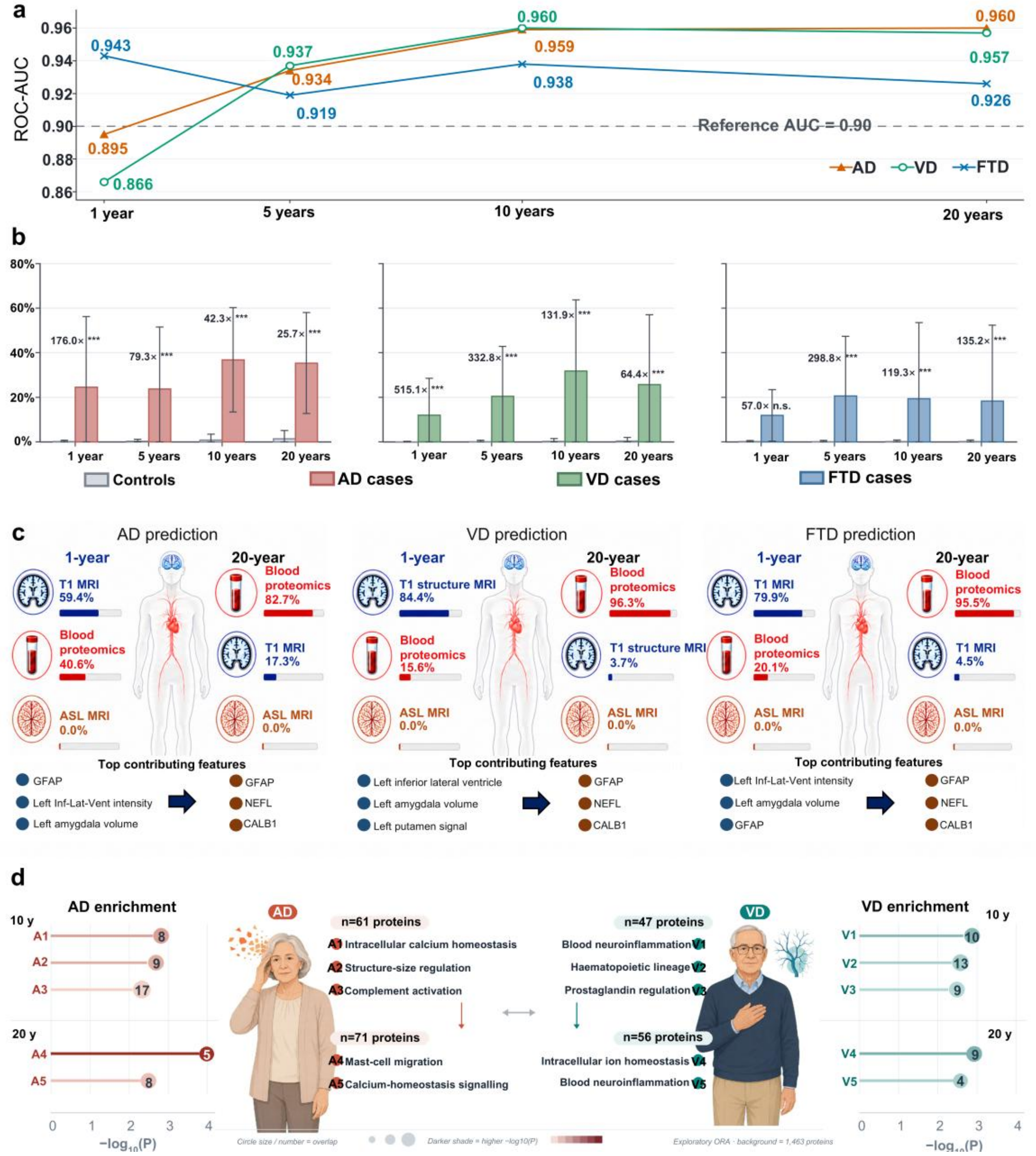

**Fig. 2 | Longitudinal dementia risk prediction reveals temporal shifts in biological determinants.** **a**, Discrimination performance of NetMoint for Alzheimer's disease (AD), vascular dementia (VD) and frontotemporal dementia (FTD) across 1-, 5-, 10- and 20-year prediction horizons, assessed by area under the receiver operating characteristic curve (ROC–AUC). **b**, Calibration and risk stratification of NetMoint-derived probabilities across prediction horizons. Predicted risks are shown for incident dementia cases and controls, with fold differences indicating enrichment among future cases. **c**, Horizon-dependent modality attribution based on normalized absolute SHAP values. The relative contributions of structural MRI, plasma proteomics and arterial spin labelling (ASL) MRI features were

assessed across dementia subtypes and prediction horizons. **d**, Biological pathway enrichment analysis of highly attributed proteomic features identified by NetMoint for AD and VD across prediction horizons. Risk estimates represent NetMoint-derived probabilities and should not be interpreted as observed cumulative incidence. SHAP, SHapley Additive exPlanations; ROC–AUC, area under the receiver operating characteristic curve.

## 3. NetMoint captures shared, subtype-specific and temporally evolving biological signatures

To determine whether the learned representations contained biologically meaningful dementia-related information, we first evaluated modality-specific representations in held-out data. Proteomic, structural MRI and cerebral haemodynamic representations discriminated participants with all-cause dementia from healthy controls with AUCs of 0.9740, 0.9532 and 0.9335, respectively. Pretraining improved these AUCs by 5.95%, 5.68% and 3.46%, respectively, relative to models trained without pretraining. SHAP attribution and age- and sex-adjusted association analyses further showed concordance between model attribution and disease-associated phenotypes. Among highly attributed features, 46 proteomic, 51 structural MRI and 38 cerebral haemodynamic measures were associated with all-cause dementia after false discovery rate correction (Supplementary Fig. 1). Prominent signals included circulating GFAP and NEFL, temporal and hippocampal structural measures, and thalamic haemodynamic alterations. The direction of case–control differences was broadly concordant with feature attribution, indicating that the model representations captured measurable dementia-related biological variation rather than solely exploiting predictive correlations.

The representations also retained information that distinguished dementia subtypes. Structural MRI provided the strongest discrimination for AD and VD, with AUCs of 0.990 and 0.974, respectively, whereas proteomic and cerebral haemodynamic representations retained substantial subtype information, including for FTD, with AUCs of 0.889 and 0.934, respectively. GFAP and NEFL were prominent proteomic features across subtypes, whereas TFF2 and BCAN showed greater prominence in AD and KLK4 and LCN2 in VD. AD-associated structural signals involved medial temporal, ventricular and cortical regions, while cerebral haemodynamic signals included arterial transit time and cerebral blood flow measures in subcortical and frontal territories. Thus, NetMoint encoded both shared dementia-related information and features with greater subtype preference.

The relative contribution of these modalities changed systematically with the prediction horizon (Fig. 2c). Structural MRI features accounted for most of the attribution at the 1-year horizon but contributed substantially less at 20 years, whereas circulating proteomic features showed the opposite pattern. At shorter horizons, the most informative features were dominated by ventricular, deep grey-matter, corpus callosum and medial temporal structural measures, together with GFAP. At longer horizons, the attribution profile became predominantly proteomic and included proteins related to neurodegeneration, inflammation and metabolism, such as GFAP, NEFL, CALB1, FCN2, PLA2G7, IGF2R, MAPT and CCL27 (Supplementary Fig. 2). Cerebral haemodynamic features were less prominent in the global attribution profile but showed consistent contributions in modality-specific analyses, particularly measures of cerebral blood flow and arterial transit time across cortical, white-matter and subcortical vascular territories. These findings indicate that the biological information used for dementia risk prediction shifts with time, from structural brain vulnerability at shorter horizons towards circulating molecular signatures of broader biological susceptibility at longer horizons.

We next examined the biological processes represented by the long-horizon proteomic signatures using exploratory over-representation analysis (Fig. 2d). AD-associated proteins at the 10-year horizon were enriched for calcium homeostasis ($P = 1.51 \times 10^{-3}$; overlap = 8), regulation of anatomical structure ($P = 2.10 \times 10^{-3}$; overlap = 9) and complement activation ($P = 4.72 \times 10^{-3}$; overlap = 17). At 20 years, immune-related processes, particularly mast-cell migration, became more prominent, while calcium-homeostasis pathways remained represented. The enrichment profile for VD was distinct, with neuroinflammatory ($P = 1.32 \times 10^{-3}$; overlap = 10), haematopoietic ($P = 2.60 \times 10^{-3}$; overlap = 13) and prostaglandin-related processes ($P = 3.25 \times 10^{-3}$; overlap = 9) prominent at intermediate horizons, followed at longer horizons by enrichment for ion homeostasis ($P = 1.12 \times 10^{-3}$; overlap = 9) alongside persistent neuroimmune signatures ($P = 2.56 \times 10^{-3}$; overlap = 4). In complementary cross-sectional analyses, AD-associated proteins were enriched for protein secretion, negative regulation of cell–matrix adhesion, negative regulation of protein binding and long-term memory, whereas VD-associated proteins showed enrichment for protein secretion, endothelial proliferation involved in sprouting angiogenesis and long-term memory. These exploratory analyses provide biological context for the model-derived molecular signatures but do not establish longitudinal pathway changes or causality.

**4. Multi-horizon risk trajectories identify distinct patterns and highly concentrated dementia risk**

Accurate risk estimates across multiple horizons enabled us to characterize how future dementia risk was distributed across individuals. We grouped the four horizon-specific predicted probabilities using a prespecified two-stage clustering framework that separated a low-risk background population from participants with distinct cross-horizon risk patterns. The resulting solutions were highly stable across bootstrap samples, with adjusted Rand indices of 0.983 for AD, 1.000 for FTD and 0.984 for VD. Most participants belonged to a background group with persistently negligible predicted risk throughout follow-up (Fig. 3a–c).

Among participants outside the background groups, risk trajectories differed substantially across dementia subtypes. AD was characterized by gradually increasing risk (29.8% of the cohort; 0.17% to 1.81%), high-increasing risk (6.5%; 0.40% to 11.43%) and a small persistently very-high-risk group (0.7%; 5.88% to 53.50%). FTD showed greater temporal heterogeneity, including a subgroup whose predicted risk declined from 0.59% at 1 year to near zero at 20 years (14.8%), alongside slowly increasing (19.3%; 0.21% to 0.62%) and persistently very-high-risk trajectories (11.24% to 67.17%). VD was characterized predominantly by a gradually increasing low-risk trajectory (26.6%; 0.04% to 1.46%) and a smaller rapidly increasing group (n = 105; 4.65% to 70.74%). Thus, dementia susceptibility followed distinct temporal patterns rather than a single monotonic trajectory, with FTD showing the greatest heterogeneity across prediction horizons.

Importantly, the highest predicted risks were concentrated within small trajectory-defined populations. The persistently very-high-risk AD group comprised only 0.7% of the cohort but reached a 20-year predicted risk of 53.50%, while the corresponding very-high-risk FTD and rapidly increasing VD groups reached 67.17% and 70.74%, respectively. To assess whether these model-derived trajectories reflected clinically meaningful heterogeneity, we examined baseline phenotypes that were not used for model training (Fig. 3d–f and Supplementary Fig. 3). Higher-risk trajectories consistently showed less favourable profiles across education, cognition, physical function and lifestyle. After multiple-testing correction, differences were observed for all 20 phenotypes examined in AD, 18 in FTD and 16 in VD, with education showing the largest overall effect and walking pace, fluid intelligence and smoking status showing consistent associations. These findings indicate that multi-horizon risk trajectories capture clinically relevant heterogeneity and identify small populations in which future dementia risk is disproportionately concentrated.

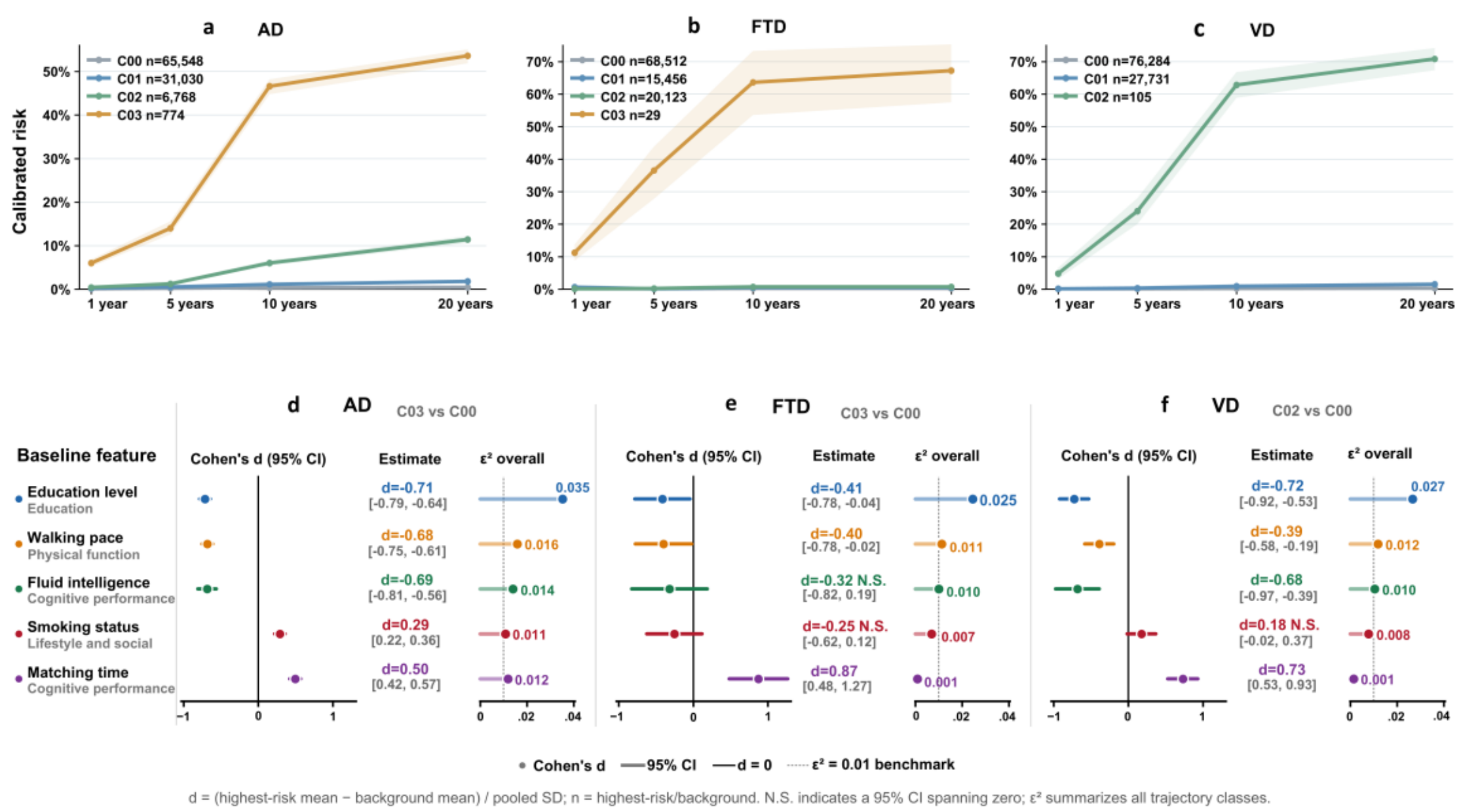

**Fig. 3 | Multi-horizon risk trajectories reveal heterogeneous patterns of dementia susceptibility.** **a-c**, Clustering of NetMoint-derived risk profiles identified distinct temporal trajectories for AD (a), FTD (b) and VD (c) across the 1-, 5-, 10- and 20-year prediction horizons. **d-f**, Associations between trajectory membership and representative baseline phenotypes not used for model training. Standardized effect sizes are shown for comparisons between the highest-risk and background trajectory groups, together with overall effect sizes across trajectory groups. Lines denote mean predicted probabilities and shaded regions indicate 95% confidence intervals.

## 5. Molecular features distinguish divergent risk trajectories within dementia subtypes

We next asked whether distinct risk trajectories persisted among individuals who ultimately developed the same dementia subtype. Subtype-specific clustering of incident cases based on their multi-horizon risk profiles, with cluster solutions selected using complementary measures of cluster separation and bootstrap stability, identified distinct trajectory-defined groups across AD, FTD and VD (Fig. 4 and Supplementary Fig. 4). AD cases comprised a stable low-to-moderate-risk group (48.1%), a group with low early risk followed by a marked increase at 10 years and subsequent plateau (43.3%), and a persistently high-risk group (8.6%). Most FTD cases showed persistently low predicted risk (80.7%), whereas smaller groups followed increasing high-risk (11.0%) or increasing very-high-risk (8.3%) trajectories. VD cases showed greater heterogeneity, comprising persistently low risk (65.6%), low early risk followed by a late increase (16.1%), low early risk followed by a sharp increase at 10 years and subsequent plateau (13.6%), and increasing very-high-risk profiles (4.8%). Thus, even within a common eventual dementia diagnosis, individuals differed substantially in the temporal pattern and magnitude of antecedent risk.

We then examined whether these trajectory differences were accompanied by molecular features that could distinguish individuals within the same dementia subtype. Among incident AD cases, TGFB1 abundance differed across trajectory groups ($q = 0.0297$), with the lowest levels in the persistently high-risk group. Among incident FTD cases, NDRG1 abundance similarly differed across trajectories ($q = 0.0454$), with the highest levels in the increasing very-high-risk group (Fig. 4d,e). The distributions of both proteins were concentrated in specific high-risk trajectories rather than following a simple monotonic gradient across groups. Thus, individuals sharing the same eventual dementia subtype could nevertheless be separated into distinct temporal risk states by baseline molecular features.

To determine whether these associations extended beyond discrete trajectory groups, we related protein abundance to predicted risk on a continuous scale after individually excluding TGFB1 or NDRG1 from model training and retraining NetMoint. In the full cohort, higher TGFB1 abundance was associated with a lower overall level of predicted AD risk across the four horizons ($\beta = -0.106$, 95% CI −0.116 to −0.096, $P = 7.28 \times 10^{-99}$; Fig. 5a,b). The association remained consistent at 1, 5, 10 and 20 years ($\beta = -0.108$, −0.104, −0.097 and −0.116, respectively), with modest variation across horizons (protein-by-horizon interaction: Wald $\chi^2 = 30.029$, df = 3, $P = 1.36 \times 10^{-6}$; BH-adjusted $P = 2.72 \times 10^{-6}$), and remained significant in sensitivity analyses using generalized estimating equations and 1% tail trimming (both $P < 0.01$). Among participants who subsequently developed AD, TGFB1 remained associated with the overall level of predicted AD risk ($\beta = -0.103$, 95% CI −0.185 to −0.021, $P = 0.014$), without evidence of horizon-dependent variation (protein-by-horizon interaction: Wald $\chi^2 = 0.918$, df = 3, $P = 0.821$).

NDRG1 showed a distinct relationship with FTD risk. Among participants who subsequently developed FTD, higher NDRG1 abundance was associated with greater predicted FTD risk ($\beta = 0.976$, 95% CI 0.468–1.483, $P = 1.66 \times 10^{-4}$), with a strong protein-by-horizon interaction (Wald $\chi^2 = 61.263$, df = 3, $P = 3.16 \times 10^{-13}$). The association increased progressively from a non-significant estimate at 1 year ($\beta = 0.343$, 95% CI −0.200 to 0.885) to significant associations at 5, 10 and 20 years ($\beta = 0.803$, 1.299 and 1.458, respectively; all $P < 0.01$). The 20-year estimate was substantially greater than the 1-year estimate ($\Delta\beta = 1.116$, 95% CI 0.805–1.426, $P = 1.85 \times 10^{-12}$), and this horizon-dependent association remained significant in sensitivity analyses. This pattern paralleled the selective elevation of NDRG1 in the increasing very-high-risk FTD trajectory.

Together, these analyses show that individuals who ultimately develop the same dementia subtype can follow distinct temporal risk trajectories with different molecular signatures. Lower TGFB1 characterized the persistently high-risk AD trajectory, whereas higher NDRG1 preferentially marked the increasing very-high-risk FTD trajectory. These associations link trajectory-defined risk heterogeneity to baseline molecular features while remaining associative rather than demonstrating biological progression or causality.

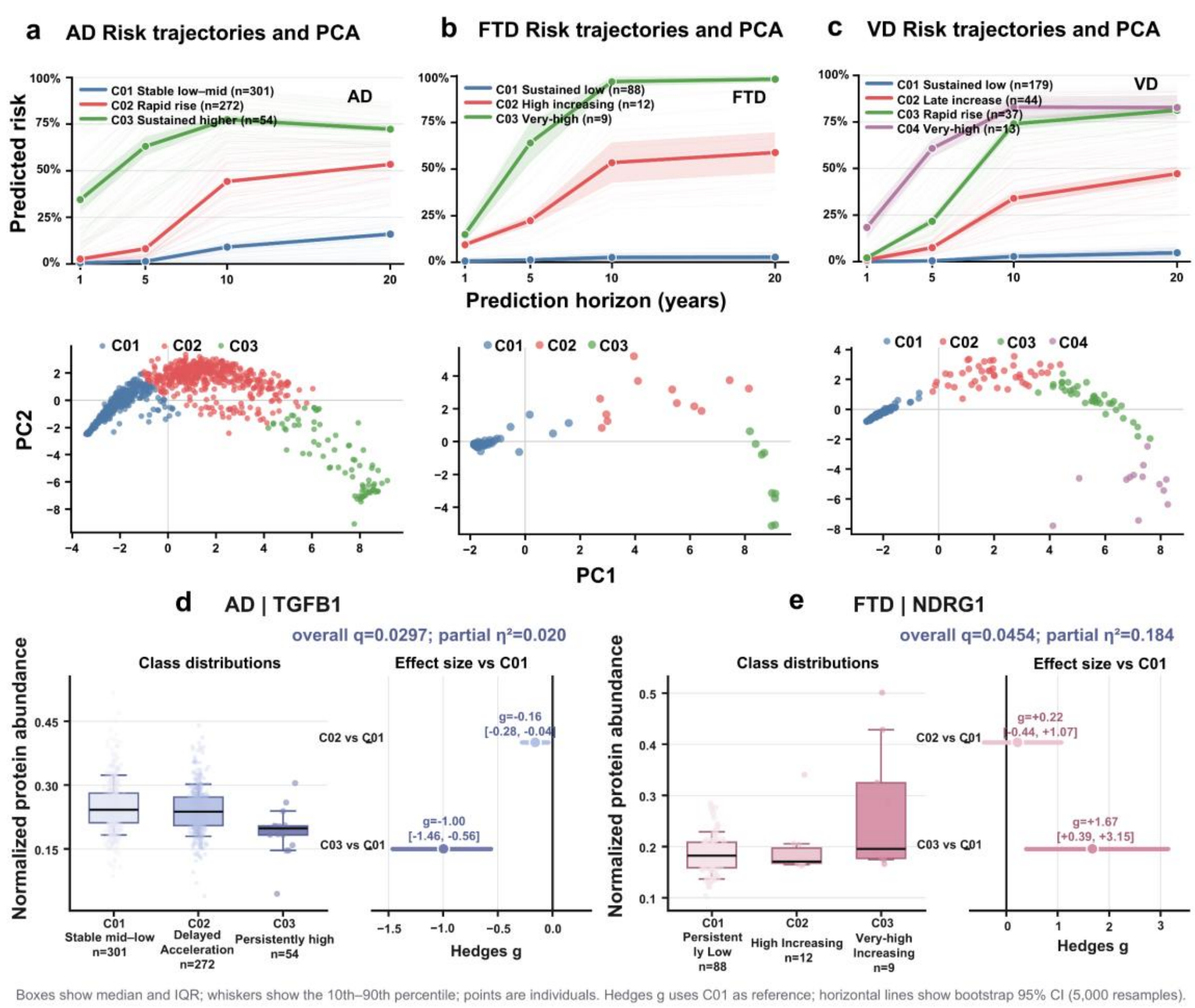

**Fig. 4 | Risk trajectories identify molecularly distinct subgroups of incident dementia. a–c**, Clustering of NetMoint-derived multi-horizon risk profiles identified distinct trajectories among incident Alzheimer's disease (AD; **a**), frontotemporal dementia (FTD; **b**) and vascular dementia (VD; **c**) cases. Within each disease subtype, clusters were ranked by ascending model-derived risk and assigned sequential labels: C01–C03 for AD and FTD, and C01–C04 for VD. Thick lines denote cluster means, faint lines represent individual risk profiles and shaded regions indicate 95% confidence intervals. Principal component projections show the separation of clusters in the feature space used for clustering. d**,e,** Associations between representative proteins and trajectory-defined subgroups. **TGFB1** abundance differed across AD trajectories (**d**), whereas **NDRG1** abundance differed across FTD trajectories (**e**), with the highest-risk trajectory groups showing the greatest molecular divergence. Boxes indicate the

median and interquartile range; whiskers denote the 10th–90th percentiles. Risk trajectories represent NetMoint-derived predicted probabilities across multiple prediction horizons and should not be interpreted as observed disease progression. CI, confidence interval; IQR, interquartile range.

### 6. Molecular features distinguish clinically similar individuals with divergent dementia risk trajectories

To determine whether multi-horizon risk profiles could capture clinically relevant heterogeneity before dementia onset, we examined individuals who subsequently developed the same dementia subtype. Despite sharing the eventual clinical diagnosis, these individuals were distributed across distinct longitudinal risk trajectories, indicating that the path to a common dementia phenotype was not uniform. This heterogeneity was particularly evident for AD and FTD, in which a subset of individuals exhibited persistently high or progressively increasing predicted risk, whereas others remained at comparatively low risk until closer to the prediction horizon (Fig. 4d and 4e).

We next asked whether these trajectory differences were accompanied by molecular features that could further distinguish individuals within the same dementia subtype. Among incident AD cases, TGFB1 abundance differed across trajectory groups ($q = 0.0297$), with the lowest levels observed in the persistently high-risk group. Among incident FTD cases, NDRG1 abundance similarly differed across trajectories ($q = 0.0454$), with the highest levels observed in the increasing very-high-risk group (Supplementary Fig. 5). The non-monotonic distribution of both proteins indicated that these associations were concentrated in specific high-risk profiles rather than reflecting a simple gradient of increasing or decreasing risk. Thus, individuals who ultimately received the same dementia diagnosis could nevertheless be distinguished by both the temporal pattern of predicted risk and their baseline molecular state.

To assess whether these molecular differences were robust beyond discrete trajectory classification, we related protein abundance to predicted risk on a continuous scale after retraining NetMoint with the corresponding protein excluded from model development. In the full cohort, higher TGFB1 abundance was associated with lower predicted AD risk across all four horizons ($\beta = -0.106$, 95% CI −0.116 to −0.096, $P = 7.28 \times 10^{-99}$; Fig. 5a, b). The association remained consistently inverse at 1, 5, 10 and 20 years ($\beta$ = −0.108, −0.104, −0.097 and −0.116, respectively), with only modest variation in effect size across horizons (protein-by-horizon interaction: Wald $\chi^2$ = 30.029, df = 3, $P = 1.36 \times 10^{-6}$; BH-adjusted $P = 2.72 \times 10^{-6}$), and was preserved in sensitivity analyses using generalized estimating equations and 1% tail trimming (both $P < 0.01$). Among participants who subsequently developed AD, the association remained significant ($\beta = -0.103$, 95% CI −0.185 to −0.021, $P = 0.014$) but did not vary by prediction horizon (protein-by-horizon interaction: Wald $\chi^2$ = 0.918, df = 3, $P = 0.821$), indicating that lower TGFB1 was associated with a higher overall level of predicted AD risk rather than with a particular temporal pattern of risk.

NDRG1 showed a different temporal relationship with FTD risk. Among participants who subsequently developed FTD, higher NDRG1 abundance was associated with greater predicted FTD risk ($\beta = 0.976$, 95% CI 0.468–1.483, $P = 1.66 \times 10^{-4}$), and the association strengthened progressively with the prediction horizon (protein-by-horizon interaction: Wald $\chi^2$ = 61.263, df = 3, $P = 3.16 \times 10^{-13}$). The association was not significant at 1 year ($\beta = 0.343$, 95% CI −0.200 to 0.885) but increased at 5, 10 and 20 years ($\beta$ = 0.803, 1.299 and 1.458, respectively; all $P < 0.01$), with the 20-year estimate substantially exceeding the 1-year estimate ($\Delta\beta = 1.116$, 95% CI 0.805–1.426, $P = 1.85 \times 10^{-12}$). This temporal pattern mirrored the selective elevation of NDRG1 in the increasing very-high-risk FTD trajectory, linking a discrete high-risk phenotype with a continuous molecular–risk relationship.

Together, these analyses extended the trajectory-based findings to continuous molecular–risk associations. Lower TGFB1 levels were associated with a higher overall level of predicted AD risk, whereas higher NDRG1 levels preferentially marked longer-term FTD risk and the increasing very-high-risk trajectory. Thus, individuals who ultimately develop the same dementia subtype may follow distinct temporal risk trajectories with different molecular signatures, suggesting that dementia risk assessment may benefit from moving beyond subtype-based diagnosis towards trajectory-resolved and biologically informed stratification.

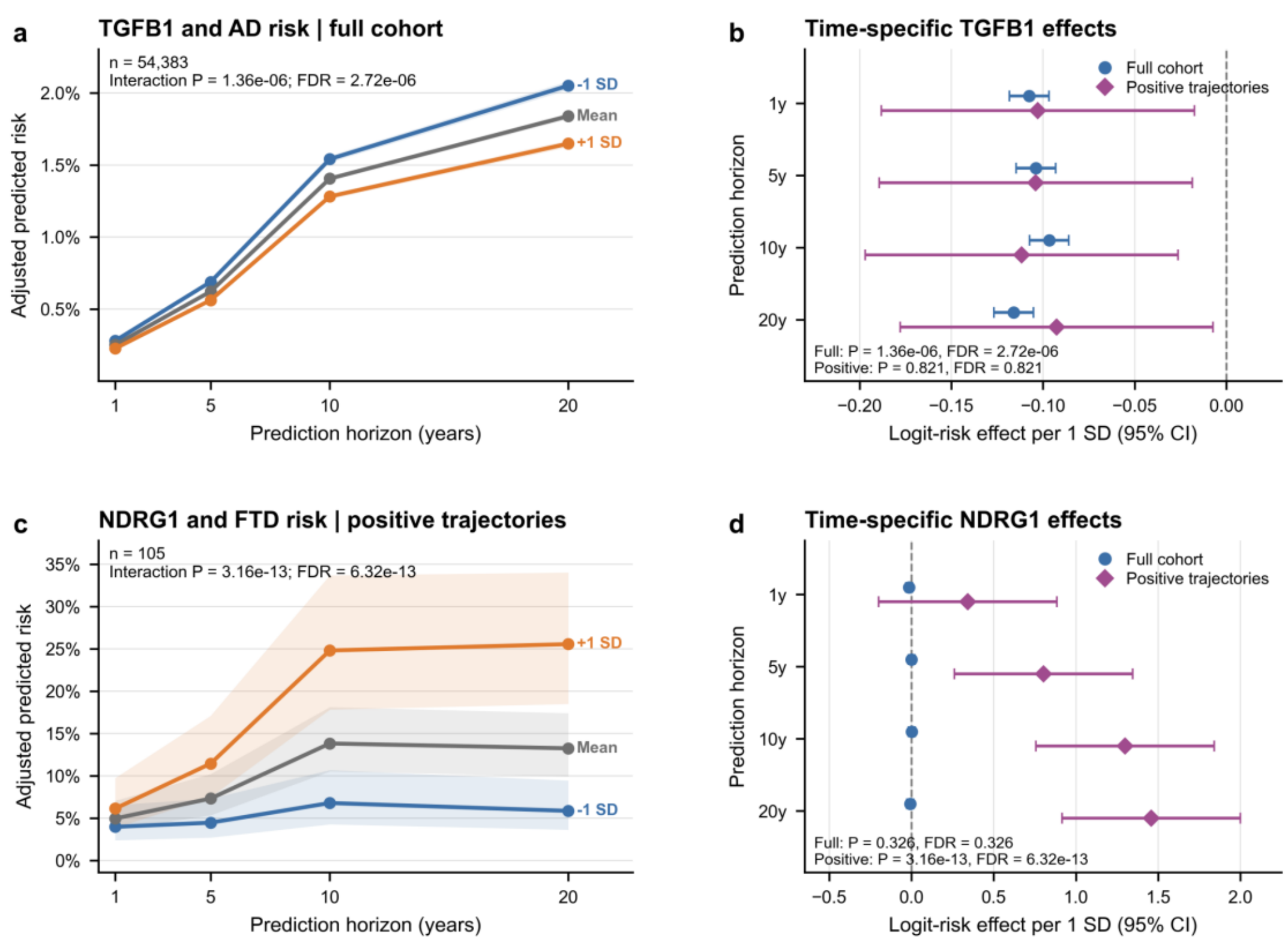

**Fig. 5 | Distinct temporal associations of TGFB1 and NDRG1 with model-derived dementia risk.** **a,c**, Adjusted marginal NetMoint-predicted risks at protein levels of −1 s.d., the mean and +1 s.d. are shown for TGFB1 and AD in the full cohort (**a**) and for NDRG1 and FTD among participants with positive FTD trajectories (**c**). Higher TGFB1 was consistently associated with lower AD risk, whereas higher NDRG1 was associated with progressively greater FTD risk at longer horizons. Lines denote adjusted predicted risks and shaded regions indicate 95% confidence intervals. **b,d**, Horizon-specific effects of TGFB1 on AD risk (**b**) and NDRG1 on FTD risk (**d**) in the full cohort and among participants classified as

positive for the corresponding subtype at any horizon. The horizontal axis represents the change in logit-transformed NetMoint-predicted risk per 1-s.d. increase in protein abundance; negative and positive values indicate associations with lower and higher predicted risk, respectively. TGFB1 showed a stable inverse association across horizons, whereas NDRG1 showed little association in the full cohort but an increasingly positive association at longer horizons among participants with positive FTD trajectories. Points denote effect estimates and error bars indicate 95% confidence intervals. Global protein-by-horizon interaction P values and false-discovery-rate-adjusted P values are shown. Risk estimates represent model-derived probabilities and not observed cumulative incidence or biological disease progression. AD, Alzheimer's disease; FTD, frontotemporal dementia; s.d., standard deviation.

## 7. External validation supports cross-cohort transport of AD risk prediction

To further assess the transportability of NetMoint, we trained the model in the Alzheimer's Disease Neuroimaging Initiative (ADNI) and evaluated it in an independent UK Biobank cohort using a harmonized multimodal feature space (Supplementary Fig. 6). Batch-proxy correction and feature matching reduced the original feature space to 138 measures shared across the two cohorts. Despite differences in cohort composition and feature availability, the ADNI-trained model retained moderate discrimination for AD risk in UK Biobank, with ROC–AUC values increasing from 0.694 at 5 years to 0.741 at 20 years (Fig. 6a). The difference in discrimination between the development and external cohorts also decreased at longer prediction horizons, indicating that the model preserved informative AD risk ranking across cohorts despite substantial feature restriction.

We next examined whether the biological signals underlying AD prediction were also reproducible across cohorts. Several proteomic associations showed concordant directions, including positive associations for IGFBP2 and myoglobin in both ADNI and UK Biobank. Structural MRI provided stronger evidence of replication: lower bilateral hippocampal, left amygdala, left accumbens and right thalamic volumes, together with higher right lateral ventricular volume, were consistently associated with AD across cohorts (Fig. 6b–e). The most prominent replicated structural signals involved the right hippocampus and right lateral ventricle. These findings nevertheless require cautious interpretation because only 27 AD cases were available in the UK Biobank structural MRI subset.

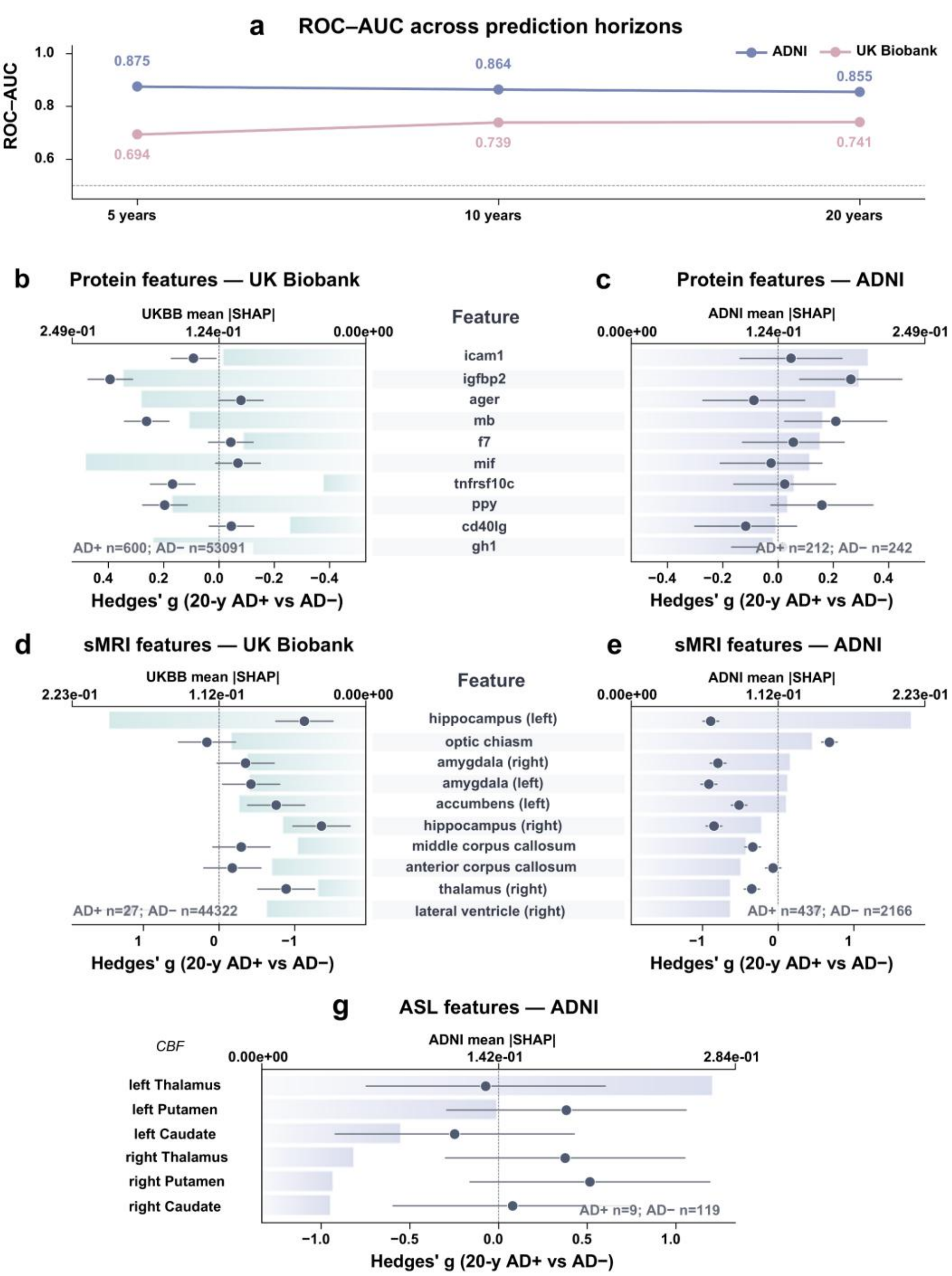

**Fig. 6 | Cross-cohort validation of ADNI-trained multimodal Alzheimer's disease risk prediction in UK Biobank.** **a**, ROC–AUC of the ADNI-trained model at the 5-, 10- and 20-year prediction horizons in ADNI and the independent UK Biobank cohort. The dashed line indicates chance-level discrimination (AUC = 0.5). **b**,**c**, Proteomic feature attribution and case–control effects at the 20-year horizon in UK Biobank (b) and ADNI (c). **d**,**e**, Corresponding structural MRI feature attribution and case–control effects in UK Biobank (d) and ADNI (e). Light bars indicate mean absolute SHAP values. Points and horizontal lines indicate Hedges' g estimates and 95% confidence intervals for AD-positive versus AD-negative participants. **g**, ASL feature attribution and case–control effects in ADNI. ASL results were exploratory because of the small number of AD-positive participants, and corresponding UK Biobank effects could not be estimated because no AD-positive cases were available in the ASL subset. Sample sizes are shown within each panel. AD, Alzheimer's disease; ASL, arterial spin labelling; sMRI, structural magnetic resonance imaging; SHAP, SHapley Additive exPlanations.

Replication of cerebral haemodynamic signals was less conclusive because of limited event numbers. Only nine AD cases were available in the ADNI arterial spin labelling (ASL) subset, and no AD cases were present in the corresponding UK Biobank subset. Although the resulting effect estimates were imprecise, thalamic cerebral blood flow emerged as a prominent AD-associated signal in ADNI, consistent with the haemodynamic findings in the primary UK Biobank analysis. Taken together, the external analyses provide evidence that NetMoint retains AD risk-discrimination across cohorts and that several of its proteomic and structural associations show concordant biological signals, while replication of haemodynamic features remains limited by sparse events.

## Discussion

Here we show that dementia susceptibility is not a uniform or steadily increasing process, but is distributed across distinct temporal trajectories that differ by dementia subtype and, importantly, among individuals who ultimately develop the same subtype. By integrating circulating proteomics, structural

MRI and cerebral haemodynamic imaging, NetMoint captured these trajectories across 1- to 20-year prediction horizons despite heterogeneous and incomplete multimodal observations. Most individuals remained at persistently low predicted risk, whereas a small proportion showed disproportionately high and sustained or rapidly increasing risk, indicating that future dementia risk is concentrated within distinct high-risk populations rather than distributed along a single continuum. The biological information supporting prediction also varied with temporal distance from diagnosis, with structural brain alterations contributing more strongly to near-term risk and circulating molecular signatures becoming increasingly informative for longer-term risk. Within incident dementia, trajectory-defined groups were further distinguished by molecular features, with lower TGFB1 marking a persistently high-risk AD trajectory and higher NDRG1 marking an increasing very-high-risk FTD trajectory. Together, these findings suggest that dementia risk assessment may be better framed as trajectory-resolved stratification, in which the timing and magnitude of future risk, rather than a single probability of disease occurrence, help distinguish individuals who may warrant different levels of surveillance and biological assessment.

*Concentrated high-risk trajectories may provide a clinically relevant dimension of risk stratification*

An important feature of the trajectory analysis was that the highest predicted risks were concentrated in relatively small subgroups rather than distributed uniformly across the population. In AD, for example, only 0.7% of the cohort followed a persistently very-high-risk trajectory, yet their predicted 20-year risk reached 53.5%. A similar pattern was observed for FTD, in which 11.24% of participants followed a persistently very-high-risk trajectory with a 20-year predicted risk of 67.17%, whereas VD included a small rapidly increasing group whose predicted risk reached 70.74%. These findings suggest that population-level discrimination and individual-level risk concentration capture complementary properties of a prediction model. A model may perform well overall while its greatest clinical relevance lies in identifying a relatively small subset in whom future risk is markedly concentrated.

This distinction may be particularly relevant as dementia prevention moves towards earlier intervention. Rather than treating risk as a continuous gradient that increases uniformly with age, trajectory-resolved profiles could identify individuals whose predicted risk is already substantially elevated over long horizons, as well as those whose risk remains low until later in the preclinical period. The clinical value of such stratification will require prospective evaluation, but the framework provides a potential basis for risk-adapted surveillance, selective longitudinal biomarker assessment and enrichment of prevention trials. Importantly, the trajectory groups were not defined solely by model inputs: higher-risk profiles were also associated with less favourable education, cognitive, physical-function and lifestyle phenotypes that were withheld from model training. This external phenotypic concordance supports the clinical relevance of the risk profiles rather than simply reflecting differences in the variables used to construct them.

The changing contribution of individual modalities across prediction horizons provides a potential biological interpretation of these temporal risk estimates. Structural MRI features were most informative for near-term prediction, with highly attributed features involving ventricular enlargement, deep grey matter, corpus callosum and medial temporal structures, consistent with measurable neuroanatomical alterations closer to clinical manifestation [28]. By contrast, longer-term prediction increasingly relied on circulating proteins, including GFAP, NEFL, CALB1, FCN2, ITGB1BP1, PLA2G7, IGF2R, MAPT and CCL27 across dementia subtypes. This pattern is compatible with the possibility that structural imaging captures relatively proximate manifestations of neurodegeneration, whereas circulating molecular signals may reflect earlier systemic or molecular susceptibility [11, 13]. Cerebral haemodynamic measures contributed less prominently to global attribution but retained subtype-relevant signals in modality-specific analyses, supporting a potential role for vascular and perfusion-related abnormalities [29, 30].

The pathway analyses further suggested that the molecular context of long-term risk was not uniform across dementia subtypes. AD-associated proteins showed increasing representation of immune-related processes at longer horizons while retaining calcium-homeostasis signals, whereas VD retained neuroinflammatory, haematopoietic and prostaglandin-related signatures alongside greater representation of ion-homeostasis processes at longer horizons. These findings are consistent with the distinct biological processes implicated in neurodegenerative and vascular dementia [31, 32]. They also illustrate why multimodal, multi-horizon modelling may provide information that cannot be recovered from a single modality or prediction window. Importantly, these attribution patterns describe how baseline measurements contribute to future-risk estimation; they do not constitute direct evidence of biomarker evolution within individuals or define discrete stages of preclinical disease.

*Temporal risk profiles reveal heterogeneity within the same dementia diagnosis*

The trajectory analysis further showed that individuals who ultimately develop the same dementia subtype do not necessarily share a common antecedent risk pattern. AD cases separated into stable low-to-moderate, sharply increasing and persistently high-risk profiles, while FTD cases included predominantly low-risk trajectories alongside smaller groups with increasing high or very-high risk. VD showed particularly diverse trajectory patterns. This heterogeneity suggests that a common clinical diagnosis can be preceded by substantially different patterns of accumulated risk, consistent with the multifactorial biology of dementia and, particularly, the contribution of vascular, metabolic and neurodegenerative processes to VD [31, 32].

The molecular analyses provided an initial demonstration that these trajectory differences are biologically distinguishable. TGFB1 and NDRG1 were not simply among the strongest population-level predictors; instead, their distributions differentiated specific high-risk trajectories. Lower TGFB1 abundance characterized the persistently high-risk AD group, whereas higher NDRG1 abundance characterized the increasing very-high-risk FTD group. The distinction was further supported by continuous analyses after excluding each protein from model development. TGFB1 remained indicative of a higher overall level of predicted AD risk, whereas NDRG1 showed an increasingly strong association with FTD risk at longer prediction horizons. Thus, the trajectory analysis identified molecular heterogeneity that could be obscured by conventional global feature ranking.

The TGFB1 finding has some directional support from previous peripheral-blood studies. A large cross-sectional plasma proteomic analysis implicated TGFB1-containing immune networks across neurodegenerative diseases and reported lower AD-associated abundance for one TGFB1-targeting aptamer [33]. Higher plasma TGF-β1 has also been associated with better hippocampal preservation in older adults [34], whereas lower concentrations have been associated with subsequent cognitive decline in adults with Down syndrome [35]. These observations are compatible with reduced circulating TGFB1 marking altered systemic immunoregulatory capacity. However, the Olink assay quantifies relative abundance of the TGFB1 proprotein, and this association should therefore not be interpreted as direct evidence of reduced active TGF-β signalling. NDRG1, by contrast, has not been established as a circulating biomarker of FTD [33]. Its established roles in hypoxic, inflammatory and endothelial cellular stress responses [36] raise the possibility that circulating NDRG1 reflects peripheral cellular stress or tissue turnover rather than an FTD-specific secreted pathway. The NDRG1 finding should therefore be regarded as a candidate horizon-dependent signal requiring orthogonal assay validation and replication in independent FTD cohorts.

*Multimodal integration addresses an important constraint on clinical implementation*

A practical strength of NetMoint is its ability to learn from partially observed multimodal phenotypes. Biological measurements are rarely available uniformly across individuals, particularly in population-based cohorts in which imaging and proteomic assessments are determined by study design, clinical indication or participant characteristics [24, 25]. By learning a shared representation from incomplete proteomic, structural MRI and cerebral haemodynamic measurements, NetMoint avoids the need to construct separate prediction models for each missing-modality pattern. This property allowed the model to retain discrimination across dementia subtypes and prediction horizons despite heterogeneous modality coverage.

However, tolerance to missing modalities does not itself remove bias arising from non-random data availability [26, 27]. The observed performance may partly reflect the underlying cohort structure and the processes determining who receives each assessment. Future evaluations should therefore characterize missingness mechanisms explicitly, examine performance across clinically relevant modality combinations and test whether learned representations remain robust in independent clinical populations. The preliminary cross-cohort analysis provides some support for transportability, but the substantial reduction in harmonized features and limited subtype-specific events highlight the need for broader validation.

Several limitations should be considered when interpreting these findings and assessing the clinical readiness of NetMoint. Although NetMoint showed strong discrimination within UK Biobank and preliminary transportability across cohorts, generalizability to independent clinical populations remains to be established. The cross-cohort analysis used a separately trained ADNI model and was restricted to 138 harmonized features shared with UK Biobank; moreover, small numbers of cases with structural MRI or ASL limited modality-specific analyses, and comparable evaluations were not possible for VD or FTD [37, 38]. A second limitation concerns the outcome definitions. Dementia subtypes were derived primarily from clinical records, were not mutually exclusive and generally lacked biomarker or neuropathological confirmation, while the relatively small numbers of subtype-specific cases, particularly FTD, limited some analyses. Third, although masked multimodal learning allows prediction from incomplete observations, it does not eliminate bias arising from non-random patterns of data availability or assessment selection. Finally, the multi-horizon risk profiles reflect differences in baseline-derived predictions rather than directly observed trajectories of biological change, and the associations of TGFB1 and NDRG1 with specific risk profiles remain exploratory. Although analyses were designed to reduce circularity, these associations do not establish causal roles or clinical utility. Future studies incorporating biomarker-confirmed diagnoses, longitudinal multimodal measurements and independent, more diverse cohorts will be essential to establish generalizability, clarify biological mechanisms and determine the clinical value of NetMoint.

Future studies should focus on external validation, dynamic risk updating and clinical translation of NetMoint. Independent cohorts with greater ancestral diversity, larger numbers of subtype-specific cases and harmonized multimodal measurements will be needed to establish generalizability across populations and missing-data patterns [39, 40]. Longitudinal integration of repeated molecular, imaging and clinical measurements may enable dynamic risk updating and better capture changing disease susceptibility [41, 42], whereas the candidate proteins identified from specific risk trajectories, including TGFB1 and NDRG1, require independent biological validation [13]. Ultimately, prospective studies should establish whether this approach improves clinical decision-making and personalized prevention. NetMoint provides a framework for moving dementia prediction beyond static risk assessment towards individualized, biologically informed and temporally structured risk stratification.

## Methods

### Study design and participants

This study was designed to develop a multimodal framework for estimating subtype-specific dementia risk across multiple prediction horizons using prospective UK Biobank (UKB; application no. 98148) data. UKB is a large population-based cohort with linked longitudinal clinical records, proteomic measurements and multimodal neuroimaging data [43]. The study integrated three complementary data modalities: circulating plasma proteomics, structural magnetic resonance imaging (sMRI) and cerebral haemodynamic features derived from arterial spin labelling (ASL).

Participants were eligible for inclusion if they had at least one available modality among proteomic, structural imaging or haemodynamic measurements, allowing the model to accommodate heterogeneous patterns of data availability. Proteomic measurements acquired at imaging/assessment instance 0 were considered the proteomic baseline, whereas neuroimaging measurements obtained at instance 2 were considered the imaging baseline. Age at each assessment-centre visit was obtained from UKB Data-Field 21003 and was used to align modality-specific measurements with individual follow-up time. For risk prediction analyses, participants with prevalent target dementia diagnoses at baseline were excluded. Individuals without sufficient follow-up information or who were censored before the corresponding prediction horizon were excluded from horizon-specific analyses. All analyses were conducted in accordance with UK Biobank material transfer agreements and ethical approval procedures.

## Dementia outcomes and prediction horizons

Dementia outcomes were defined for three major clinical subtypes: Alzheimer's disease (AD), vascular dementia (VD) and frontotemporal dementia (FTD). Incident diagnoses were identified from linked inpatient hospital records using ICD-10 diagnostic codes recorded in UK Biobank Data-Field 41202. No additional diagnostic sources were incorporated into the primary outcome definition. For each dementia subtype, cumulative risk labels were generated for four prediction horizons (1, 5, 10 and 20 years) relative to the prediction origin. Participants were assigned a positive outcome if the corresponding dementia diagnosis was recorded within the specified horizon and were otherwise classified as negative. Because dementia subtypes may overlap clinically and pathologically, diagnoses were not treated as mutually exclusive. Accordingly, NetMoint was formulated as a multilabel prediction framework, in which subtype-specific cumulative risks were estimated independently across prediction horizons.

## Blood proteomic data acquisition and preprocessing

The proteomic input consisted of 1,463 circulating proteins quantified in 55,298 participants from the UK Biobank [15]. Protein abundance data underwent quality control and upstream batch-effect correction before inclusion in the present study. Plate and well identifiers were obtained from UK Biobank Data-Fields 30901 and 30902, respectively, and were used to characterize potential technical variation. Protein distributions before and after batch correction are shown in Supplementary Fig. 7. Following preprocessing, all protein features were normalized before model development. Because batch correction had already been performed during the upstream data processing pipeline, additional batch adjustment was not applied during NetMoint training to avoid redundant correction and potential distortion of biologically relevant variation.

**Modality availability and missing modality handling**

During preprocessing, a modality was considered available when fewer than 10% of modality-specific features were missing; otherwise, the modality was considered unavailable for that participant. Missing modalities were represented using zero-filled tensors solely as computational placeholders to maintain consistent batch dimensions during model training. Binary modality-availability masks were incorporated into the fusion module to prevent these placeholder values from contributing to feature aggregation. Therefore, zero-filled tensors did not represent biological measurements or imputed observations.

In the final analytical dataset, 53,199 participants had proteomic data only, 42,648 had structural MRI (sMRI) data only, and 5,528 had cerebral haemodynamic data from arterial spin labelling (ASL) only. Multimodal availability included 1,120 participants with both proteomics and sMRI, 64 with proteomics and ASL, and 646 with sMRI and ASL. All three modalities were available in 915 participants. Overall, all 104,120 participants contributed at least one available modality. Modality availability profiles were recalculated and verified against the original source datasets before model development.

This masking strategy enabled NetMoint to integrate heterogeneous modality combinations without requiring complete-case data. However, it did not assume that missingness occurred at random and could not eliminate potential selection effects associated with disease status, participation patterns or imaging acquisition. Therefore, interpretability analyses were additionally performed in the full cohort, the independent test set and the subset of participants with complete availability of all three modalities.

## NetMoint multi-omics architecture

NetMoint consisted of three major components: modality-specific variational autoencoders (VAEs) [44], a masked precision-weighted product-of-experts (PoE) [45] fusion module, and a horizon- and subtype-conditioned risk prediction head. The final implementation included 2,878,542 trainable parameters.

Proteomics, structural MRI (sMRI) and arterial spin labelling (ASL) features were processed using independent modality-specific encoders. Each encoder comprised sequential hidden layers with 128, 256 and 128 units. Each hidden layer consisted of a linear transformation followed by batch normalization, rectified linear unit (ReLU) activation, dropout (0.2) and a residual block. The encoder outputs were used to parameterize the mean and log variance of a 256-dimensional latent representation. A symmetric multilayer perceptron decoder was used to reconstruct the original modality input through a 256–128–256 architecture followed by projection to the corresponding input dimension. Encoder log variances were constrained to predefined ranges before being used for latent fusion.

For each participant i, modality m availability was incorporated into the PoE fusion through an explicit availability mask.

$$\tau_{im} = \exp(-\ell_{im})$$

The fused posterior was defined as:

$$\tau_i = \sum\nolimits_m a_{im}\,\tau_{im}, \quad \mu_i = \frac{\sum_m a_{im}\,\tau_{im}\mu_{im}}{\tau_i + 10^{-8}}, \quad \ell_i = -\log(\tau_i + 10^{-8})$$

No expert contribution was assigned to unavailable modalities, and an additional prior expert was not introduced.

$$z_i = \mu_i + \exp(0.5\ell_i) \odot \varepsilon, \quad \varepsilon \sim \mathrm{N}(0, I), \quad z_i \in [-5,5]$$

The risk head learned separate 16-dimensional embeddings for each prediction horizon and disease subtype. Concatenated horizon and disease embeddings generated a 256-dimensional FiLM scaling term and offset [46].

The scaling term was constrained by a tanh transformation. The conditioned latent representation was:

$$z_{\text{cond}} = (1 + \gamma) \odot z + \beta$$

The conditioned latent representation was further transformed through a horizon–subtype projection and subsequently reduced to 128 dimensions before entering two residual blocks [47]. Each residual block incorporated sigmoid gating, GELU activation, dropout (0.3) and batch normalization. The final prediction layer generated a single logit for each horizon–subtype combination. The same conditional risk head was applied across four prediction horizons and three dementia subtypes, resulting in 12 horizon-specific risk estimates.

The modality-attention module was implemented as an auxiliary diagnostic component. It generated modality-level outputs for diagnostic interpretation but was not involved in construction of the PoE latent representation or the downstream risk prediction pathway. Therefore, modality-attention outputs were not interpreted as quantitative contributions to model predictions.

**Model training and optimization**

Each modality-specific autoencoder was initially pretrained independently using data from the corresponding modality. The pretrained components were subsequently integrated into the multimodal framework and jointly optimized using the following composite objective:

$$L = 0.1L_{\text{recon}} + 0.001L_{\text{KL}} + 5L_{\text{BCE}}$$

The reconstruction loss was calculated as the mean squared error between reconstructed and observed features within available modalities. Reconstruction errors were first averaged across observed features within each modality and subsequently aggregated across the modalities available for each participant. The Kullback–Leibler (KL) divergence term represented the analytical divergence between the fused Gaussian posterior distribution and a standard normal prior. The binary cross-entropy loss was computed from logits for available horizon-specific dementia outcome labels across the 12 prediction tasks, with missing labels excluded from loss calculation.

Model optimization was performed using the Adam optimizer [48] with an initial learning rate of $1 \times 10^{-3}$ and a batch size of 64 for a maximum of 20 epochs. Model performance was monitored on the validation set during training. When the validation objective failed to improve for 10 consecutive epochs, the learning rate was reduced by a factor of 0.1 using the ReduceLROnPlateau scheduler, with a minimum learning rate of $1 \times 10^{-6}$.

**Probability calibration and predictive performance**

Following model training, temperature scaling was performed separately for each dementia subtype and prediction horizon using the validation set [48]. The temperature parameter (T) was optimized by minimizing the negative log-likelihood with bounded L-BFGS-B optimization over the range of 0.01–100. The optimized calibration parameters were fixed and applied to the logits of the independent test set. Classification thresholds were determined exclusively using validation-set predictions and subsequently applied unchanged to the test set.

Primary model evaluation was conducted exclusively in the independent test set. Predictive performance was assessed separately for each dementia subtype and prediction horizon using the area under the receiver operating characteristic curve (AUROC), area under the precision–recall curve (AUPRC), F1 score, precision, recall, sensitivity and specificity. Predictions generated in the full cohort were used only for descriptive risk characterization, interpretability analyses and hypothesis-generating analyses, and were not used to estimate model generalizability.

**ADNI cohort and outcome ascertainment**

The Alzheimer’s Disease Neuroimaging Initiative (ADNI) served as the external evaluation cohort. We included plasma proteomics, regional structural MRI measures and arterial spin labelling (ASL) measures of cerebral blood flow. The datasets contained 1,062 proteomic, 11,329 structural MRI and 346 ASL subject–visit observations. Their union contained 11,363 observations. Each observation represented a subject visit rather than a unique participant. ADNI was conducted with local institutional review-board approval and written informed consent.

Diagnostic status was reconstructed from the longitudinal ADNI diagnosis summary. The diagnosis date was used as the label anchor when available; otherwise, the corresponding modality date was used. Current status was defined by the latest eligible diagnosis on or before the anchor. AD required DIAGNOSIS = 3 and AD attribution, defined as DXAD = 1 in ADNI-1 or DXDDUE = 1 in later phases. Cognitively normal and mild cognitive impairment observations were grouped as non-AD to match the model labels. Observations with non-AD dementia, competing dementia aetiologies or unresolved diagnoses were excluded.

Incident AD was assessed within 1, 5, 10 and 20 years after the label anchor. A horizon was positive when AD occurred within the corresponding interval. Baseline AD and the first incident AD event were propagated to all subsequent horizons. Strict negative labels required documented follow-up through the corresponding horizon without AD or competing dementia. Insufficient follow-up was coded as missing.

For the complete binary matrix required by the model, unresolved horizons were completed using a prespecified permissive rule. Missing intervals before a known AD event were assigned 0. Later unresolved horizons were assigned 0 when the last observed horizon was negative and no AD event was recorded. Current non-AD observations without longitudinal follow-up were also assigned 0. These values represent extrapolated non-events and do not indicate documented AD-free follow-up for 20 years. The strict labels were retained for sensitivity analysis.

Feature names were harmonized with the derivation-cohort dictionary. Only high-confidence mappings were retained, yielding 85 plasma proteins, 47 structural MRI measures and six ASL measures. Proteomic features underwent reference-distribution normalization and outcome-independent correction for available technical proxies. Plate and well information was unavailable; this correction therefore did not provide plate-level batch harmonization. Structural MRI features underwent k-nearest-neighbour imputation and reference-distribution normalization. ASL features were normalized within acquisition-protocol strata. Outcome labels were not used during preprocessing.

**Cross-cohort model training and evaluation**

The derivation cohort included only observations with explicit AD or non-AD labels. Records with vascular dementia, frontotemporal dementia, other dementia, uncertain aetiology or discordant diagnoses were excluded. The resulting datasets contained 53,691 proteomic, 44,349 structural MRI, 6,233 ASL and 102,447 multimodal observations.

The derivation and ADNI matrices had identical feature names, ordering and dimensions. Only the 138 high-confidence shared features retained observed values. Cohort-specific and unmeasured features were set to 0 in both cohorts. These values encoded unavailable measurements rather than biological zeros.

Derivation-cohort participants were assigned to training, validation and internal-test partitions in an 80:10:10 ratio using stratified random splitting with a fixed seed of 42. All observations from the same participant were assigned to one partition. ADNI data were not used for model fitting, preprocessing estimation, feature selection, hyperparameter selection, checkpoint selection or threshold optimization. The model and preprocessing transformations were frozen before external evaluation.

The cross-cohort model used the NetMoint product-of-experts variational autoencoder. Available modalities were combined through precision-weighted latent fusion. A binary head estimated current AD versus non-AD status in single-modality analyses. A conditional temporal head estimated AD risk within 1, 5, 10 and 20 years in the multimodal analysis. Training minimized reconstruction loss, Kullback–Leibler divergence and masked binary cross-entropy. Optimization used AdamW with an initial learning rate of $1 \times 10^{-4}$, weight decay of $1 \times 10^{-3}$ and a batch size of 64.

The primary performance measure was the area under the receiver-operating-characteristic curve. Secondary measures were the area under the precision–recall curve and Brier score. Threshold-dependent metrics were calculated using the Youden-index threshold selected in the derivation validation partition. This threshold was fixed before internal testing and ADNI evaluation.

**Feature attribution and modality contribution**

Feature-level attribution was performed using GradientExplainer within the SHapley Additive exPlanations (SHAP) framework [49]. Explanations were generated separately for each dementia subtype (AD, VD and FTD) and prediction horizon (1, 5, 10 and 20 years). The primary attribution analysis was conducted in complete cases with all three available modalities from the independent test set. The SHAP background distribution consisted of 200 randomly sampled complete cases from the training set. During SHAP estimation, modalities were perturbed jointly along the same background trajectories to preserve the observed dependence structure between modalities.

Model parameters were kept fixed during interpretation. To minimize stochastic variation from the variational latent representation, the posterior mean of the fused PoE latent distribution was used as input to the risk prediction head rather than randomly sampled latent vectors. For each prediction task, SHAP values were estimated using three independent Monte Carlo seeds with 200 expected-gradient samples per seed, and the resulting attribution values were averaged across seeds. Participant-level bootstrap resampling with 2,000 repetitions was performed to estimate 95% confidence intervals for feature contributions and feature rankings.

SHAP values were calculated for each input feature and participant within the explanation cohort. Global feature importance was quantified as the mean absolute SHAP value across participants. These attribution values were interpreted as model-derived contributions to risk prediction rather than causal effects or biological effect estimates.

**Biological characterization of disease-associated features**

To characterize the biological features associated with subtype-specific dementia risk and their temporal evolution, we performed downstream feature-level analyses based on model-derived feature contributions. Candidate features were defined by combining the global top 120 features ranked by mean absolute SHAP values for each disease–horizon prediction task with prespecified modality-specific candidates. This resulted in 215, 263 and 259 candidate features for AD, VD and FTD, respectively.

For each disease subtype and prediction horizon, participants were classified according to whether they developed the corresponding subtype within the predefined prediction window. Control participants were required to remain free of all three dementia subtypes within the same horizon. Individuals diagnosed with another dementia subtype but not the target subtype were excluded from the corresponding control group to minimize contamination from subtype overlap.

Feature distributions were evaluated after adjustment for age and sex, consistent with the analyses presented in the Results. Standardized between-group differences were quantified using Hedges' g with 95% confidence intervals estimated using the Hedges–Olkin analytical approximation. Two-sided Mann–Whitney U tests were used to assess distributional differences between cases and controls. Multiple testing was controlled using the Benjamini–Hochberg procedure within each disease subtype–horizon feature family. Features satisfying both an FDR-adjusted q value $<0.05$ and an absolute effect size $|g|\geq 0.20$ were considered statistically supported.

To ensure reliable estimation of effect sizes, formal statistical inference was restricted to comparisons containing at least 10 cases. Comparisons with 3–9 cases were considered descriptive only, whereas analyses with fewer than three cases were not performed. For temporal analyses, only comparisons with at least 20 participants in each group were included. For selected high-priority features, robustness of effect size estimates was further assessed using stratified non-parametric bootstrap resampling of cases and controls with 2,000 iterations.

**Temporal trajectory analysis of feature effects**

To investigate whether biomarker associations emerged at different stages before dementia diagnosis, participants were categorized according to the time interval between baseline assessment and first subtype diagnosis: ≤1 year, 1–5 years, 5–10 years and 10–20 years. Monotonic temporal trends in feature distributions were assessed using Spearman correlation, followed by Benjamini–Hochberg correction within each dementia subtype.

For the top 20 features within each modality, inverse-variance weighted regression models were used to estimate temporal changes in standardized effect sizes (g) and effect magnitude (|g|) as a function of $\log_2$-transformed prediction horizon. Statistical significance of temporal slopes was evaluated using 5,000 parametric bootstrap samples under the null hypothesis of a constant effect over time.

Features were categorized according to their temporal behavior. Features showing a significant increase or decrease in |g| over time were classified as effect-increasing or effect-decreasing, respectively. Features with stable associations were defined as those with an absolute effect-size range ≤ 0.15

across horizons and no evidence of temporal trend ($P \geq 0.10$). Late-emerging features were identified when the 20-year effect size showed a standardized residual ≥1.96 relative to the weighted 1–10-year trajectory, together with a significant association at 20 years ($P < 0.05$), absence of significant effects at earlier horizons ($P \geq 0.05$), and a large late effect ($|g| \geq 0.50$). All remaining temporal patterns were classified as mixed or uncertain. This trajectory classification was considered exploratory.

**Subtype-specific heterogeneity analysis**

To examine whether identified features reflected subtype-specific biological processes rather than shared dementia-related alterations, heterogeneity analyses were restricted to participants with a single dementia subtype diagnosis. Feature distributions across AD, VD and FTD groups were compared using the Kruskal–Wallis test, followed by pairwise Mann–Whitney U tests with Benjamini–Hochberg correction. These analyses were designed to evaluate differences in feature distributions among clinical subtypes and were not interpreted as formal interaction tests of effect-size differences.

**Functional enrichment analysis**

To explore biological mechanisms underlying high-contribution molecular features, exploratory over-representation analysis (ORA) was performed on proteins with high model contribution for AD and VD at 10- and 20-year prediction horizons. The background set consisted of the 1,463 proteins quantified in the UK Biobank proteomic dataset.

Foreground protein sets included 61 proteins for AD at 10 years, 71 proteins for AD at 20 years, 47 proteins for VD at 10 years and 56 proteins for VD at 20 years. Enrichment analyses were performed using Gene Ontology biological processes [50], KEGG pathways[51], WikiPathways [52], MSigDB Hallmark gene sets [53] and disease-associated gene sets. Protein identifiers unavailable for specific databases were excluded, resulting in database-specific effective background sizes.

Enrichment significance was assessed using hypergeometric testing with Benjamini–Hochberg correction performed separately within each database and foreground set. Because multiple enrichment databases were analyzed independently, q values were not interpreted as representing global false discovery control across databases. Therefore, this analysis was considered exploratory, and both nominal P values and database-specific FDR-adjusted values were reported.

**Definition of multihorizon dementia risk profiles**

Temperature-calibrated probabilities generated by NetMoint for 1-, 5-, 10- and 20-year prediction horizons were integrated to define a participant-level multihorizon risk profile for each dementia subtype. These probabilities were considered model-derived risk representations capturing the predicted distribution of future disease risk across multiple time horizons and were not interpreted as longitudinal repeated measurements or estimates of cumulative incidence.

To identify participants with consistently low predicted risk, we first defined a low-risk background group (C00) using three complementary summary measures: the maximum predicted probability across horizons, the time-normalized trapezoidal area under the risk curve and overall risk amplitude. Initial thresholds were set at the 75th percentile of the corresponding subtype-specific distributions. When the resulting background group exceeded 82% of the cohort, the threshold was relaxed to the 70th percentile to preserve sufficient heterogeneity for downstream clustering.

**Identification of model-derived risk phenotypes**

Risk trajectory clustering was performed among participants outside the low-risk background group. For each dementia subtype, clustering features were constructed to capture both absolute risk levels and temporal dynamics of predicted risk. These features included raw predicted probabilities, square-root transformed probabilities, adjacent horizon differences, interval-normalized slopes, total risk change, early and late risk changes, normalized area under the risk curve, mean risk, maximum risk, risk range and log-transformed risk values:

$$x_{\log} = \log_{10}(p + 10^{-6})$$

All clustering features were standardized using subtype-specific Z-score normalization before clustering.

K-means clustering with k-means++ [54] initialization was applied with candidate cluster numbers ranging from 2 to 6. The optimal number of clusters was determined using a composite criterion integrating the silhouette coefficient (35%) [55], adjusted Rand index [56] from repeated subsampling (35%), logarithmically transformed Calinski–Harabasz index (15%) and inverse Davies–Bouldin index (15%) [57]. To reduce instability caused by very small clusters, a penalty was applied to solutions containing clusters smaller than the larger of 50 participants or 1% of the clustering population.

For model selection, clustering was performed using up to 16,000 non-background participants with a fixed random seed (20260614). The numbers of initializations for model selection, subsampling and final clustering were 25, 8 and 50, respectively, with a maximum of 800 iterations. Cluster labels were interpreted as model-derived risk patterns and did not represent clinical stages or disease progression categories.

**Sensitivity analysis among predicted-positive participants**

To assess the robustness of risk phenotypes among individuals with elevated predicted risk, an independent clustering analysis was performed among participants classified as positive for the corresponding dementia subtype at any prediction horizon. In this analysis, clustering features were restricted to interpretable risk trajectory descriptors, including raw risk, adjacent differences, temporal slope, total change, area under the risk curve, mean risk and maximum risk. Candidate cluster numbers ranged from 2 to 6 and were evaluated using 100 subsamples containing 80% of participants.

The numbers of initializations for cluster selection, subsampling and final fitting were 50, 10 and 100, respectively. The maximum number of iterations was increased to 1,000, with a fixed random seed (20260609). A penalty was applied to solutions containing clusters smaller than the larger of five participants or 2% of the sample. When multiple solutions showed comparable performance, the smaller number of clusters was preferred if the composite score was within 0.03 of the optimum and the adjusted Rand index exceeded 0.55.

Following clustering, identified risk groups were characterized using 20 baseline phenotypic variables that were not used during model training or clustering. Continuous and ordinal variables were compared across clusters using Kruskal–Wallis tests. To ensure reliable estimation of group differences, clustering groups containing fewer than 10 participants were excluded, and analyses were restricted to subtype analyses with at least 30 total participants. Effect sizes were quantified using epsilon-squared statistics. Multiple testing across the 20 phenotypic variables was controlled using the Benjamini–Hochberg procedure within each dementia subtype.

**Identification of omics features associated with risk phenotypes**

To identify molecular features associated with model-derived dementia risk phenotypes, we compared omics measurements across risk strata within each dementia subtype and modality. Feature-level differences among clusters were assessed using one-way analysis of variance. Analyses were restricted to comparisons with at least 10 participants overall and at least three participants in each cluster. Effect sizes were quantified using partial eta-squared, and multiple testing was controlled using the Benjamini–Hochberg procedure [58] within each subtype–modality analysis family.

Risk clusters were ordered according to their mean area under the predicted risk curve across horizons, representing increasing model-derived risk levels. Monotonic changes in molecular features across ordered risk strata were assessed using Spearman correlation followed by BH correction. Pairwise standardized differences relative to the lowest-risk cluster (C01) were quantified using Hedges' g and considered descriptive. Based on these analyses, the AD-associated TGFB1 and FTD-associated NDRG1 protein signals were selected for subsequent temporal modelling. ASL-derived associations for FTD were considered exploratory owing to limited sample availability.

**Time-dependent protein–risk trajectory modelling**

To investigate whether selected proteins showed horizon-dependent associations with predicted dementia risk, we performed longitudinal trajectory analyses using the multihorizon risk outputs generated by NetMoint. Analyses were conducted both in the full cohort and among participants predicted positive for the corresponding dementia subtype at any prediction horizon. The full-cohort analyses included 54,383 participants for both TGFB1–AD and NDRG1–FTD analyses. Positive-sample analyses included 600 participants with AD risk and 105 participants with FTD risk, respectively.

Protein measurements were batch-corrected during upstream UK Biobank preprocessing and subsequently normalized before modelling. Therefore, protein batch effects were not additionally adjusted in the primary analyses. Age and sex were included as covariates. Risk clusters were used only for descriptive visualization and were not included in regression models.

For each analysis, the target protein was standardized within the corresponding disease-specific analysis population. Predicted risks at the four horizons were constrained to avoid numerical instability and transformed to the logit scale before conversion from wide to long format：

$$p \in [10^{-6}, 1 - 10^{-6}], \quad \mathrm{logit}(p) = \log\left(\frac{p}{1-p}\right)$$

Linear mixed-effects models were fitted separately for each protein–disease pair, with participant-specific random intercepts accounting for repeated horizon-level measurements. The full model included the standardized protein level, prediction horizon indicators and protein-by-horizon interaction terms：

$$\mathrm{logit}(\mathrm{risk}_{it}) = \beta_0 + \beta_P \mathrm{protein}_{z,i} + \sum_t \beta_t I_t + \sum_t \beta_{P\times t} \left(\mathrm{protein}_{z,i} \times I_t\right) + b_i + \varepsilon_{it}$$

The 1-year prediction horizon was used as the reference category. A reduced model excluding protein-by-horizon interaction terms was fitted to estimate the overall association between protein level and predicted risk. Time-dependent effects were evaluated by comparing the full and reduced models using likelihood-ratio tests and a three-degree-of-freedom Wald test.

Within each analysis population, BH correction was applied to the two prespecified interaction tests evaluating TGFB1–AD and NDRG1–FTD temporal effects.

For each prediction horizon, results were reported as the logit-scale effect size per 1-SD increase in protein level, corresponding 95% confidence interval, P value and odds ratio. Model fit was assessed using Akaike information criterion, Bayesian information criterion, marginal $R^2$ and conditional $R^2$ [59]. Adjusted predicted risks and 95% confidence intervals were estimated at protein levels of −1, 0 and +1 SD using Gauss–Hermite quadrature to marginalize over the random-effect distribution.

**Sensitivity analyses and protein contribution assessment**

Sensitivity analyses were performed using participant-clustered robust standard errors and after excluding participants within the extreme 1% tails of protein distributions. Additional analyses restricted the cohort to participants predicted positive for the corresponding dementia subtype and used 1,000 participant-level bootstrap replicates with a fixed random seed (20260724). The bootstrap procedure used an influence-function approximation rather than complete model refitting for each replicate.

Model assumptions were evaluated using residual-versus-fitted plots and normal Q–Q plots. Because the outcome variable represented NetMoint-generated dementia risk rather than observed clinical incidence, these analyses were interpreted as associations between protein levels and the temporal pattern of model-derived risk estimates. They were not interpreted as causal relationships between protein expression and disease progression.

To determine whether TGFB1 and NDRG1 contributed directly to model predictions, NetMoint models were retrained after removing TGFB1 or NDRG1 from the corresponding proteomic input space. Predictive performance and risk trajectory differences from protein-deletion models were evaluated separately.

**Classification validation, statistical analysis, and reproducibility**

To further evaluate the generalizability of the learned multimodal representation, the same NetMoint latent embedding was applied to additional classification tasks, including binary all-cause dementia classification and multilabel classification of AD, VD and FTD. Classification models operated on the 256-dimensional fused posterior mean generated by the pretrained NetMoint encoder. Separate multilayer perceptron classifiers were trained for each downstream task using binary cross-entropy loss.

Statistical associations between model predictions and clinical or molecular features were evaluated in prespecified analysis sets. Continuous variables were adjusted for age and sex by linear regression before downstream analyses. Associations between continuous features and binary outcomes were assessed using partial point-biserial correlations based on the age- and sex-adjusted residuals. Case-control differences were estimated using linear regression models of the form:

$$\text{feature} \sim \text{group} + \text{age} + \text{sex}$$

Unless otherwise specified, statistical significance was assessed using two-sided tests. Empirical P values were estimated using 10,000 permutation tests where appropriate, and confidence intervals were obtained by participant-level bootstrap resampling. Multiple comparisons were controlled using the Benjamini–Hochberg procedure within each prespecified family of hypotheses. Effect sizes and corresponding 95% confidence intervals were reported together with adjusted P values.

Model development was performed in Python (v3.13.9). Statistical analyses were performed using NumPy (v2.3.5), pandas (v2.3.3), SciPy (v1.16.3) and statsmodels (v0.14.5). Data visualization was generated using matplotlib (v3.10.6). All analyses were conducted using fixed random seeds to ensure reproducibility.

# Acknowledgements

This research has been conducted using the UK Biobank Resource under Application Number [98148]. Data collection and sharing for this project were supported by the Alzheimer’s Disease Neuroimaging Initiative. ADNI data were disseminated by the Laboratory for Neuro Imaging at the University of Southern California.

This work is supported by the Key Project (Youth Program, Category B，2026JC-YXQN-271), National Key Research and Development Program of China (2025YFE0215700), Shaanxi Provincial Department of Science and Technology，Key Research and Development Project of the Shaanxi Provincial Health Science and Technology Innovation Capacity Enhancement Program (2025YF-08) and Fundamental Research Funds for the Central Universities – Young Innovative Research Team (xtr052025007).